\documentclass[pra,twocolumn,showpacs,preprintnumbers,amsmath,amssymb,superscriptaddress]{revtex4}

\usepackage{color}
\usepackage{graphicx}
\usepackage{dcolumn}
\usepackage{bm}

\def\tvt{two vs. two}
\def\ovt{one vs. two}
\def\<{\langle}
\def\>{\rangle}

\begin{document}

\title{Conjectured strong complementary-correlations tradeoff}

\author{Andrzej Grudka}

\affiliation{Faculty of Physics, Adam Mickiewicz University, 61-614 Pozna\'{n}, Poland}

\author{Micha{\l} Horodecki}

\affiliation{Institute for Theoretical Physics and Astrophysics,
University of Gda{\'n}sk, 80-952 Gda{\'n}sk, Poland}

\author{Pawe{\l} Horodecki}

\affiliation{Faculty of Technical Physics and Applied Mathematics,
 Gda{\'n}sk University of Technology, 80-952 Gda{\'n}sk, Poland}

\affiliation{National Quantum Information Centre in Gda{\'n}sk, 81-824 Sopot, Poland}

\author{Ryszard Horodecki}

\affiliation{Institute for Theoretical Physics and Astrophysics,
University of Gda{\'n}sk, 80-952 Gda{\'n}sk, Poland}

\affiliation{National Quantum Information Centre in Gda{\'n}sk, 81-824 Sopot, Poland}

\author{Waldemar K{\l}obus}

\affiliation{Faculty of Physics, Adam Mickiewicz University, 61-614 Pozna\'{n}, Poland}

\author{{\L}ukasz Pankowski}

\affiliation{Institute for Theoretical Physics and Astrophysics,
University of Gda{\'n}sk, 80-952 Gda{\'n}sk, Poland}

\date{\today}

\begin{abstract}
We conjecture new uncertainty relations which restrict
correlations between results of measurements performed by two separated parties on a shared quantum state. The first uncertainty relation bounds the sum of two mutual informations when one party measures a single observable and the other party measures one of two observables. The uncertainty relation does not follow from Maassen-Uffink uncertainty relation and is much stronger than Hall uncertainty relation derived from the latter. The second uncertainty relation bounds the sum of two mutual informations when each party measures one of two observables.
We provide numerical evidence for validity of conjectured uncertainty relations and
prove them for large classes of states and observables.
\end{abstract}

\maketitle

\section{Introduction}
The uncertainty relations impose
fundamental limitations on our ability to simultaneously predict the outcomes
of measurements of different observables. They are widely known in the form given by Robertson \cite{PhysRev.34.163} which relates variances of two operators with their commutator by  inequality:
\begin{eqnarray} \label{eq:1}
\Delta A \Delta B \geq \frac{1}{2}|\langle \psi|[A,B]|\psi \rangle|.
\end{eqnarray}
A special example of such uncertainty relations is Heisenberg uncertainty relation for position and momentum measurements which states that $\Delta x \Delta p \geq \hbar/2$.
However, for many operators the right hand side of (\ref{eq:1}) depends on a state $|\psi \rangle$ and can equal to $0$, although both variances on the left hand side of (\ref{eq:1}) cannot simultaneously equal to $0$.

Several authors recognized that one can express uncertainty relations in terms of entropies (see \cite{Wehner1, Bialynicki1} for review and \cite{PhysRevLett.103.020402, Berta1, PhysRevLett.106.110506, PhysRevLett.108.210405, PhysRevA.86.042315, Frank1, Roga1, Hasse1} for recent developments). In particular  Maassen and Uffink \cite{PhysRevLett.60.1103} inspired by the work of Deutsch \cite{PhysRevLett.50.631} derived the following entropic uncertainty relation \footnote{All logarithms in this paper are of base $2$.}
\begin{eqnarray} \label{eq:2}
S(B^{(1)})+S(B^{(2)}) \geq -\log a.
\end{eqnarray}
Here $S(B^{(s)})$  is entropy of measurement outcomes when a measurement of observable $B^{(s)}$ is performed on a state $\rho$  and $a=\max_{i,j} |\langle b^{(1)}_i|b^{(2)}_j\rangle|^2$ is the square of maximum overlap between eigenvectors $|b^{(1)}_i\rangle$ of the observable $B^{(1)}$ and eigenvectors $|b^{(2)}_j\rangle$ of the observable $B^{(2)}$. This uncertainty relation does not suffer from the previous criticism because the right hand side of (\ref{eq:2}) does not depend on $\rho$.

The result of Maassen and Uffink was extended by Hall \cite{PhysRevLett.74.3307} to derive a bound on accessible information about a quantum system represented by an ensemble of states. Let us suppose that Alice prepares a state $\rho_i$ with probability $p_i$ and Bob performs a measurement on it of an observable $B^{(1)}$ or $B^{(2)}$. Hall uncertainty principle states that the sum of two accessible informations satisfies the inequality
\begin{eqnarray}\label{eq:3}
I(B^{(1)}|{\cal{E}})+I(B^{(2)}|{\cal{E}})\leq 2\log d + \log a,
\end{eqnarray}
where $d$ is dimension of Bob's Hilbert space. Here $I(B^{(s)}|{\cal{E}})=S(B^{(s)})_{\rho} - \sum_{i} p_{i} S(B^{(s)})_{\rho_i}$ is accessible information about the ensemble of states ${\cal E}$ when a measurement of the observable $B^{(s)}$ is performed on it and $\rho=\sum_ip_i\rho_i$.

For some reason this latter direction was not further developed, even though it involves a central
quantity for communication, which is mutual information. At the same time, one can see that
Hall inequalities are actually very far from being tight, unlike the original entropic inequalities.
Secondly, once we deal with mutual information, it is tempting to consider two subsystems.
This is the feature of the  newest uncertainty principle conjectured in \cite{PhysRevLett.103.020402} and proved in \cite{Berta1}
involving conditional entropy. However, once there are two systems in the play, it is natural
to ask about the principle which is symmetric with respect to the subsystems.
None of the existing uncertainty principles has this feature.

In this paper we aim to overcome these two drawbacks. First, we propose an uncertainty relations
based on mutual information, that are stronger than Hall's one.
Second, we introduce an uncertainty relation of new type, which being not yet fully symmetric with respect to
two subsystems, exhibits symmetry of the following sort:
it involves measurements of two observables on each of two subsystems.
We provide numerical evidence for validity of conjectured relations, and
prove them for large classes of states and observables.

More specifically, we consider two parties -- Alice and Bob -- who share a quantum state and assume that Alice performs a measurement on her part of the state in some basis and Bob performs measurement on his part of the state in one of two bases. We will be interested in accessible correlations between measurements outcomes when Alice and Bob measure one pair of their observables rather than in accessible information about one party's measurement. We will derive an uncertainty relation which bounds the sum of two mutual informations -- the first one between the results of Alice's measurement and the results of Bob's first measurement and the second one between the results of Alice's measurement and the results of Bob's second measurement. Although we proved our uncertainty relation only for certain states, we suppose that it holds in general.
This relation called here {\it mutual uncertainty relation} does not follow from Maassen-Uffink uncertainty relation and is much stronger than Hall uncertainty relation derived from the latter. As a special case we present much stronger bound on accessible information about certain ensembles of states $\cal{E}$ when measurements of certain observables $B^{(s)}$ are performed on it.
Moreover, we derive mutual uncertainty relation in the case where Alice and Bob share a maximally entangled state and  each party performs a measurement in one of two bases.

The paper is organized as follows. In Section II we formulate uncertainty relation for a case when Alice measures one observable and Bob measures one of two observables. In Section III we formulate uncertainty for a case when both parties measure one of two observables. In Section IV we present analytical results for certain states and obsrvables. Finally in Section V we conclude.

\section{Mutual uncertainty relation for one vs. two observables}
Suppose that Alice performs a measurement in a basis ${\cal A}=\{|a_{k}\rangle \}$ and Bob performs a measurement in one of two bases ${\cal B}^{(s)}=\{|b_{j}^{(s)}\rangle \}$. We conjecture that the following uncertainty relation for the sum of two mutual informations -- the first one between Alice's results of measurement and Bob's results of measurement when he performs a measurement in the {\it first} basis and the second one between Alice's results of measurement and Bob's results of measurement when he performs a measurement in the {\it second} basis -- holds:
\begin{eqnarray}\label{eq:4}
I(A:B^{(1)}) + I(A:B^{(2)}) \leq \log d + \log c,
\end{eqnarray}
where $d$ is dimension of each party's Hilbert space and $c$ is the sum of $d$ largest coefficients $c_{ij}$ with
\begin{eqnarray}\label{eq:5}
c_{ij} = |\langle b_i^{(1)}|b_j^{(2)}\rangle|^2.
\end{eqnarray}
We have tested this uncertainty relation numerically for dimension of each party's subsystem up to $d=4$ and have found no violation (Details of numerical calculations are given in Appendix.).
Moreover, we prove in Theorem 1 that if after Alice's measurement the state is diagonal in a basis $|a_{k}\rangle \otimes |b_{j}^{(1)}\rangle$ then the uncertainty relation (\ref{eq:4}) holds.
However,  we were not able to prove it in general case.

Let us now compare our mutual uncertainty relation with Hall's original one.
Suppose that Alice and Bob share a state $\rho_{AB}=\sum_{i=0}^{d-1}p_i|i\rangle \langle i|\otimes |i\rangle \langle i|$. Alice performs a measurement in the basis ${\cal A}=\{|i\rangle, i=0,...,d-1\}$ while Bob performs a measurement in the basis ${\cal B}^{(1)}=\{|i\rangle, i=0,..., d-1 \}$ or ${\cal B}^{(2)}=\{|0\rangle, |\tilde{j}\rangle, \tilde{j}=1,..., d-1 \}$, where $|\tilde{j}\rangle=\frac{1}{\sqrt{d-1}}\sum_{k=1}^{d-1} \exp({\frac{2 \pi i \tilde{j}k}{d-1}})|k\rangle$. In such a case $c=2$ and, although two bases in which Bob performs the measurement have a common eigenvector, the sum of mutual informations is bounded by $\log{d}+1$. In contrast, Hall uncertainty relation gives trivial bound $2\log d$.

Hall uncertainty relation follows directly from Maassen-Uffink uncertainty relation. Hence, another possible improvement to Hall uncertainty relation could be obtained by strengthening Maassen-Uffink uncertainty relation. To see if it is possible let us write Maassen-Uffink uncertainty relation in the following way:
\begin{eqnarray} \label{eq:6a}
S(A)+S(B) \geq \min H_\infty(\{c_{ij}\}),
\end{eqnarray}
where $H_{\infty}(\{c_{ij}\})$ is min-entropy of a row or a column of a bistochastic matrix (\ref{eq:5}) and minimum is taken over all rows and columns of this matrix.
The simplest generalization of Maassen-Uffink uncertainty relation can be obtained by replacing in (\ref{eq:6a}) min-entropy $H_{\infty} (\{c_{ij}\})$ by Renyi entropy $H_{\alpha} (\{c_{ij}\})$ with $\alpha < \infty$, where $H_{\alpha}(\{p_i\})=\frac{1}{1-\alpha}\log \sum_i p_i^{\alpha}$, i.e. we can take the uncertainty relation in the form
\begin{eqnarray} \label{eq:7a}
S(A)+S(B) \geq \min \{\min_i H_\alpha(\{c_{ij}\}), \min_j H_\alpha(\{c_{ij}\})\},
\end{eqnarray}
where $H_\alpha(\{c_{ij}\})$ is Renyi entropy of a row (or a column) of a bistochasic matrix (\ref{eq:5}) and $\alpha$ is some constant to be determined. We recall that min-entropy is obtained as a limiting case of Renyi entropy
for $\alpha\to \infty$ and that Renyi entropies satisfy the inequality $H_{\alpha}(\{p_i\}) \geq H_{\beta}(\{p_i\})$ for $\alpha < \beta$.
We have checked numerically if such strengthening of Maassen-Uffink uncertainty relation is possible and found a strong evidence that the uncertainty relation (\ref{eq:7a}) does not hold in general for $\alpha<\infty$.  More presicely when we increase dimension of the system, it is violated for larger $\alpha$.

\section{Mutual uncertainty relation for \tvt\ observables}
Suppose that Alice performs a measurement in one of two bases ${\cal A}^{(s)}=\{|a_{k}^{(s)}\rangle \}$ and Bob performs a measurement in one of two bases ${\cal B}^{(s)}=\{|b_{j}^{(s)}\rangle \}$. We conjecture that the following mutual uncertainty relation for the sum of two mutual informations -- the first one between Alice's and Bob's results of measurements when both parties perform the measurements in the {\it first} bases and the second one between Alice's and Bob's results of measurements when both parties perform the measurements in the {\it second} bases -- holds:
\begin{eqnarray}
I(A^{(1)}:B^{(1)})+I(A^{(2)}:B^{(2)}) \leq 2 \log d + \log c',
\label{eq:weak}
\end{eqnarray}
where $d$ is dimension of each party's Hilbert space and $c'=\max_V \max_{i,j} |\langle b_{i}^{(1)}|VU^TV^{\dagger}|b_{j}^{(2)} \rangle|^{2}$. Here $U$ is chosen in such a way that relation
\begin{eqnarray}
U^{\dagger}|a_{k}^{(2)}\rangle=|a_{k}^{(1)}\rangle
\end{eqnarray} is satisfied for all $k$ and $V$ is unitary operator.

We note that the coefficient $c'$ in the above uncertainty relation is analogous to the coefficient $a$ in Maassen-Uffink uncertainty relation. We have tested this uncertainty relation numerically for dimensions of each party's subsystem up to $d=3$ and have found no violation.  Moreover we prove in Theorem 3 that the uncertainty relation (\ref{eq:weak}) holds if the parties perform the measurements on the maximally entangled state.

We have also found an exotic form of an uncertainty relation, which
is for a while numerically confirmed. Namely,
we have tested numerically the inequality
\begin{eqnarray}\label{cpp}
I(A^{(1)}:B^{(1)}) + I(A^{(2)}:B^{(2)}) \leq \log c'' - 2\log d,
\end{eqnarray}
with
\begin{eqnarray}
c'' = \sum_{ijkl} c_{ijkl},
\end{eqnarray}
where
\begin{eqnarray}
c_{ijkl} = \frac{|\langle a_i^{(1)}|a_j^{(2)}\rangle|^p}{|\langle b_k^{(1)}|b_l^{(2)}\rangle|^p}.
\end{eqnarray}
for $d$ up to $16$ and  $p=\frac{1}{2}$ and have found no violation. Unfortunately, the minimal value of RHS gets closer to $2 \log d$ when the dimension $d$ increases. Moreover, for some choices of observables the RHS becomes singular.

For convenience in Table I we summarize all different coefficients for our uncertainty relations.

\begin{table*}[t]\label{tab}
\begin{equation*}
\begin{array}{lcc}
\hline \hline
\textrm{Coefficient}  & \textrm{defined as...}     &  \textrm{pertaining to relation...}  \\ \hline
c   & \sum_{d\,\,\, \textrm{largest}} |\<b^{(1)}_i|b^{(2)}_j \>|^2                     &  I(A:B^{(1)}) + I(A:B^{(2)}) \leq \log d + \log c  \\
c'  & \max_V \max_{i,j} |\<b^{(1)}_i|V U^T V^\dagger|b^{(2)}_j \>|^2                   &  I(A^{(1)}:B^{(1)})+I(A^{(2)}:B^{(2)}) \leq 2 \log d + \log c' \\
c'' & \sum_{i,j,k,l} \frac{|\<a^{(1)}_i|a^{(2)}_j \>|^{1/2}}{|\<b^{(1)}_k|b^{(2)}_l \>|^{1/2}} &  I(A^{(1)}:B^{(1)}) + I(A^{(2)}:B^{(2)}) \leq \log c'' - 2\log d \\ 
\hline \hline
\end{array}
\end{equation*}
\caption{Different coefficients appearing in uncertainty relations \eqref{eq:4}, \eqref{eq:weak}, and \eqref{cpp}. See the main text for more details.}
\end{table*}

\section{Analytical results}
Now, we prove the uncertainty relations for some states and observables. We begin with the uncertainty relations for \ovt\ observables. We shall consider an auxiliary uncertainty relation,
which is not true in general. However, in the lemma below we will show that
it holds for some states and observables, and then we will argue that
this implies validity of our relation \eqref{eq:4}. In contrast, the latter uncertainty relation
is conjectured to hold in general.

{\bf Lemma 1.} Suppose that Alice and Bob share a state $\rho_{AB}$. Alice performs a measurement in a basis ${\cal A}=\{|a_{k}\rangle \}$ corresponding to one-dimensional projectors $\{P_{k}=|a_{k}\rangle \langle a_{k}|\}$ and Bob performs a measurement in one of two bases ${\cal B}^{(s)}=\{|b_{j}^{(s)}\rangle \}$  corresponding to one-dimensional projectors $\{Q_{j}^{(s)}=|b_{j}^{(s)}\rangle \langle b_{j}^{(s)}|\}$, where the index $s=1,2$ corresponds to two bases. If after Alice's measurement the state is diagonal in a basis $|a_{k}\rangle \otimes |b_{i}^{(1)}\rangle$, i.e. it is of the form
\begin{equation}\label{eq:6}
\rho_{AB^{(1)}}=\sum_{ki} p_{ki} P_{k} \otimes Q_{i}^{(1)},
\end{equation}
then the following uncertainty relation holds:
\begin{eqnarray}
& I(A:B^{(1)})+I(A:B^{(2)}) \leq \nonumber\\
& \log d+ \log \sum_{ij}  c_{ij}^{2},
\label{eq:mun_4_H2}
\end{eqnarray}
where $d$ is dimension of each party's Hilbert space and
\begin{equation}
c_{ij}=|\langle b_{i}^{(1)}|b_{j}^{(2)}\rangle |^{2}.
\end{equation}

{\it Proof.}
If after Alice's measurement Bob performs a measurement in the basis ${\cal B}^{(1)}=|b_{j}^{(1)}\rangle $ then the state does not change. On the other hand, if Bob performs a measurement in the basis ${\cal B}^{(2)}=|b_{j}^{(2)}\rangle$ then the state takes the form
\begin{eqnarray}
& \rho_{AB} = \sum_{ijk} p_{ki} P_{k} \otimes Q_{j}^{(2)} Q_{i}^{(1)} Q_{j}^{(2)} \equiv  \nonumber\\
& \sum_{ijk} p_{ki} c_{ij} P_{k} \otimes Q_{j}^{(2)}.
\label{uncorrelated}
\end{eqnarray}
For simplicity, let us first assume that after Alice's measurement the state is of the form
\begin{equation}
\rho_{AB}=\sum_i p_i P_{i} \otimes Q_{i}^{(1)},
\label{correlated}
\end{equation}
which remains the same after Bob's measurement in the first basis. If Bob performs a measurement in the second basis  then the state becomes
\begin{equation}
\rho_{AB}= \sum_{ij}p_i c_{ij} P_{i} \otimes Q_{j}^{(2)}.
\end{equation}
Let us calculate the sum of two mutual informations. We have
\begin{eqnarray}\label{derivation1}
& I(A:B^{(1)}) + I(A:B^{(2)}) = S(A) + I(A:B^{(2)})= \nonumber \\
& S(A) + S(B^{(2)}) - S(B^{(2)}|A) = \nonumber\\
&  S(A) + S(B^{(2)})  + \sum_{i} p_i \sum_j c_{ij} \log c_{ij} \leq \nonumber\\
& S(A) + S(B^{(2)}) +  \sum_{i} p_i \log ( \sum_j c_{ij}^{2}) = \nonumber\\
& S(B^{(2)}) +  \sum_{i} p_i \log( \frac{ \sum_j c_{ij}^{2}}{p_i} ) \leq \nonumber\\
& \log d + \log( \sum_{ij} c_{ij}^{2}).
\end{eqnarray}
In the fourth and sixth lines we used the concavity of the logarithm. This concludes the proof for the state (\ref{correlated}).

For the initial state after Alice's measurement of the more general form  (\ref{uncorrelated})
it is enough to observe that this state can be obtained from the correlated one (\ref{correlated})
by applying a local channel on Alice's side which: (i) does not increase mutual informations $I(A:B^{(1)})$ and $I(A:B^{(2)})$; (ii) commutes with the second Bob's measurement (in fact with both of them, but the first is irrelevant); (iii) does not change the entropy $S(B^{(2)})$.

Because $\sum_{i}p_i\sum_j c_{ij} \log c_{ij} \leq \log \max_{i,j} c_{ij}$ we can replace $\sum_{i}p_i\sum_j c_{ij} \log c_{ij}$ in the third line of (\ref{derivation1}) by $\log \max_{i,j} c_{ij}$ and obtain the following inequality
\begin{eqnarray}
& I(A:B^{(1)}) + I(A:B^{(2)}) \leq \nonumber\\
& 2 \log d + \log \max_{i,j} c_{ij}.
\end{eqnarray}
This is a special case of Hall uncertainty relation. $\blacksquare$

For the above states and observables we can immediately prove the uncertainty relation (\ref{eq:4}).

{\bf Theorem 1.} Under the assumptions of Lemma 1 the uncertainty relation (\ref{eq:4}) holds.

{\it Proof.} Note that for a given $i$ $\{ c_{ij} \}$ is a probability distribution. Hence, we have $\sum_j c_{ij}^{2} \leq \max_{j} c_{ij} $. Taking the sum over $i$ we obtain $\sum_{ij} c_{ij}^{2} \leq \sum_i \max_{j} c_{ij} \leq \sum c_{ij}$, where the last sum is over $d$ largest coefficients $c_{ij}$. $\blacksquare$

In the following example we show that the uncertainty relation (\ref{eq:mun_4_H2}) is not valid in general.

{\bf Example.} Suppose that  Alice's and Bob's subsystems are three dimensional one. Alice performs a measurement in the basis
$\{ |a_1\rangle=(1,0,0), |a_2\rangle=(0,1,0), |a_3\rangle=(0,0,1)\}$ and Bob performs a measurement either in the basis
$|b_1^{(1)}\rangle=(1,0,0), |b_2^{(1)}\rangle=(0,1,0), |b_3^{(1)}\rangle=(0,0,1)$
or in the basis
$\{|b_1^{(2)}\rangle=(\frac{1}{\sqrt{2}},\frac{1}{\sqrt{2}},0),
|b_2^{(2)}\rangle=(\frac{1}{2},-\frac{1}{2}, \frac{1}{\sqrt{2}}), |b_3^{(2)}\rangle=(-\frac{1}{2},\frac{1}{2},\frac{1}{\sqrt{2}})\}$. Hence, the matrix of coefficients $c_{ij}$ takes the following form
\begin{eqnarray}
c_{ij}=|\langle b_{i}^{(1)}|b_{j}^{(2)} \rangle|^{2}=\left[\begin{array}{clcr} \frac{1}{2} & \frac{1}{4} & \frac{1}{4}\\
\frac{1}{2} & \frac{1}{4} & \frac{1}{4}\\
0 & \frac{1}{2} & \frac{1}{2} \end{array}\right].
\end{eqnarray}
Let us try to bound the sum of two mutual informations as in (\ref{eq:mun_4_H2}). We obtain
\begin{eqnarray}
& I(A:B^{(1)})+I(A:B^{(2)})=\nonumber \\
& \log 3 + \log \frac{5}{4}=\log \frac{15}{4}<2.
\end{eqnarray}
Now consider the following state
\begin{eqnarray}
|\Psi\rangle_{AB}=\frac{1}{\sqrt{2}}(|a_1\rangle \otimes |b_3^{(1)}\rangle+|a_2\rangle \otimes |b_1^{(2)}\rangle).
\end{eqnarray}
The sum of two mutual informations is equal to $2$ and hence it violates the bound (\ref{eq:mun_4_H2}).
We stress, however that the uncertainty relation (\ref{eq:4}) still holds in this case.

Now, we turn our attention to the uncertainty relations for \tvt\ observables. We assume that Alice and Bob share maximally entangled states and both Alice and Bob can choose one of two measurements. Our results are given in two theorems. In Theorem 2 we derive a state dependent uncertainty relation (i.e. the RHS of the uncertainty relation depends on both the choice of observables and the choice a of maximally entangled state) and in Theorem 3 we derive a state independent uncertainty relation (i.e. the RHS of the uncertainty relation depends only on the choice of observables and is valid for an arbitrary maximally entangled state).

{\bf Theorem 2.} Suppose that Alice and Bob share a maximally entangled state $|\Phi\rangle_{AB}$ which is related to the maximally entangled state $|\Phi^+\rangle_{AB} = \frac{1}{\sqrt{d}}\sum_i |i\rangle \otimes |i\rangle$ by the equation $|\Phi\rangle_{AB}=I\otimes V|\Phi^+\rangle_{AB}$, where $V$ is a unitary operation acting on Bob's subsystem. Alice performs a measurement in one of two bases ${\cal A}^{(s)}=\{|a_{k}^{(s)}\rangle \}$ corresponding to one-dimensional projectors $\{P_{k}^{(s)}=|a_{k}^{(s)}\rangle \langle a_{k}^{(s)}|\}$ and Bob performs a measurement in one of two bases ${\cal B}^{(s)}=\{|b_{j}^{(s)}\rangle \}$  corresponding to one-dimensional projectors $\{Q_{j}^{(s)}=|b_{j}^{(s)}\rangle \langle b_{j}^{(s)}|\}$. The following uncertainty relation holds:
\begin{eqnarray}
I(A^{(1)}:B^{(1)})+I(A^{(2)}:B^{(2)}) \leq 2 \log d + \log \tilde{c}',
\label{eq:theorem-3'}
\end{eqnarray}
where $d$ is dimension of each party's Hilbert space and $\tilde{c}'=\max_{i,j} |\langle b_{i}^{(1)}|VU^TV^{\dagger}|b_{j}^{(2)} \rangle|^{2}$ with $U$ chosen in such a way that relation
\begin{eqnarray}
U^{\dagger}|a_{k}^{(2)}\rangle=|a_{k}^{(1)}\rangle
\end{eqnarray}
is satisfied for all $k$.

In order to prove Theorem 2 we will need two lemmas, which are given below.

{\bf Lemma 2.} Mutual information between Alice and Bob calculated on a state
\begin{eqnarray}
\sum_{k,j}  \langle \Phi^+|_{AB} P_{k}^{(2)}\otimes Q_{j}^{(2)} |\Phi^+\rangle_{AB} P_{k}^{(2)}\otimes Q_{j}^{(2)}
\end{eqnarray}
is equal to mutual information between Alice and Bob calculated on a state
\begin{eqnarray}
& \sum_{k,j}  \langle \Phi^+|_{AB} U^{\dagger} P_{k}^{(2)} U \otimes U^T Q_{j}^{(2)} U^{*} |\Phi^+\rangle_{AB} \times \nonumber\\
& U^{\dagger}P_{k}^{(2)}U\otimes U^TQ_{j}^{(2)}U^{*},
\end{eqnarray}
where $|\Phi^+\rangle_{AB} = \frac{1}{\sqrt{d}}\sum_i |i\rangle \otimes |i\rangle$.

{\it Proof.} We prove it by showing that the former state can be transformed to the latter one by local unitary operations (which do not change mutual information). Indeed, we have
\begin{eqnarray}
& \sum_{k,j}  \langle \Phi^+|_{AB} P_{k}^{(2)}\otimes Q_{j}^{(2)} |\Phi^+\rangle_{AB} \times \nonumber\\
& U^{\dagger}P_{k}^{(2)}U\otimes U^TQ_{j}^{(2)}U^{*}=\nonumber \\
&  \sum_{k,j}  \langle \Phi^+|_{AB} U^{\dagger} P_{k}^{(2)} U \otimes U^T Q_{j}^{(2)} U^{*} |\Phi^+\rangle_{AB} \times \nonumber\\
& U^{\dagger}P_{k}^{(2)}U\otimes U^TQ_{j}^{(2)}U^{*},
\end{eqnarray}
where we used the identity  $U\otimes U^*|\Phi^+\rangle_{AB}=|\Phi^+\rangle_{AB}$. $\blacksquare$

{\bf Lemma 3.} Suppose that Alice and Bob share the maximally entangled state $|\Phi^+\rangle_{AB}$. Alice performs a measurement in one of two bases ${\cal A}^{(s)}=\{|a_{k}^{(s)}\rangle \}$ corresponding to one-dimensional projectors $\{P_{k}^{(s)}=|a_{k}^{(s)}\rangle \langle a_{k}^{(s)}|\}$ and Bob performs a measurement in one of two bases ${\cal B}^{(s)}=\{|b_{j}^{(s)}\rangle \}$  corresponding to one-dimensional projectors $\{Q_{j}^{(s)}=|b_{j}^{(s)}\rangle \langle b_{j}^{(s)}|\}$. The following uncertainty relation holds:
\begin{eqnarray}
I(A^{(1)}:B^{(1)})+I(A^{(2)}:B^{(2)}) \leq 2 \log d + \log \tilde{c},
\end{eqnarray}
where $d$ is dimension of each party's Hilbert space and $\tilde{c}=\max_{i,j} |\langle b_{i}^{(1)}|U^T|b_{j}^{(2)} \rangle|^{2}$ with $U$ chosen in such a way that relation
\begin{eqnarray}
U^{\dagger}|a_{k}^{(2)}\rangle=|a_{k}^{(1)}\rangle
\end{eqnarray}
is satisfied for all $k$.

{\it Proof.} We use Lemma 2 with $U$ chosen as above and replace the second Alice's measurement in the basis ${\cal A}^{(2)}$ given by the projectors $P_{k}^{(2)}$ by the first Alice's measurement in the basis ${\cal A}^{(1)}$ given by the projectors $P_{k}^{(1)}$ and the second Bob's measurement given in the basis ${\cal B}^{(2)}$ by the projectors $Q_{j}^{(2)}$ by the measurement in the basis ${\cal B}^{(2)}_{U}$given by the projectors $U^TQ_{j}^{(2)}U^{*}$.
We write the sum of two mutual informations in the following way
\begin{eqnarray}
\label{eq:33}
& I(A^{(1)}:B^{(1)})+I(A^{(2)}:B^{(2)})=\nonumber\\
& I(A^{(1)}:B^{(1)})+I(A^{(1)}:B^{(2)}_{U})=\nonumber\\
& S(B^{(1)})-S(B^{(1)}|A^{(1)})+S(B^{(2)}_{U})-S(B^{(2)}_{U}|A^{(1)}) \leq \nonumber\\
& 2 \log d-[S(B^{(1)}|A^{(1)})+S(B^{(2)}_{U}|A^{(1)})].
\end{eqnarray}
Let us now bound the terms in the square bracket. We have
\begin{eqnarray}
\label{eq:34}
& S(B^{(1)}|A^{(1)})+S(B^{(2)}_{U}|A^{(1)})=\nonumber \\
& \sum_{k}\frac{1}{d}[S(B^{(1)})_{\rho_{Bk}}+S(B^{(2)}_{U})_{\rho_{Bk}}],
\end{eqnarray}
where
\begin{eqnarray}
\rho_{Bk}=d\text{Tr}_{A}(P_{k}^{(1)} \otimes I |\Phi^+\rangle \langle \Phi^+|_{AB}).
\end{eqnarray}
Maassen-Uffink uncertainty relation states that
\begin{eqnarray}
\label{eq:MU}
S(B^{(1)})_{\rho_{Bk}}+S(B^{(2)}_{U})_{\rho_{Bk}} \geq - \log \tilde{c}.
\end{eqnarray}
Substituting (\ref{eq:MU}) into (\ref{eq:34}) and then substituting the result into (\ref{eq:33}) we obtain
\begin{eqnarray}
& I(A^{(1)}:B^{(1)})+I(A^{(2)}:B^{(2)})\leq \nonumber\\
& 2 \log d+\log \tilde{c}.
\end{eqnarray}
$\blacksquare$

We are now ready to prove Theorem 2.

{\it Proof.} We note that Alice's and Bob's measurements given by the projectors $P_{k}^{(s)}$ and $Q_{j}^{(s)}$performed on the maximally entangled state $|\Phi\rangle_{AB}=I\otimes V|\Phi^+\rangle_{AB}$ are equivalent to measurements given by the projectors $P_{k}^{(s)}$ and $V^{\dagger}Q_{j}^{(s)}V$performed on the maximally entangled state $|\Phi^+\rangle_{AB}$. Then, from Lemma 3 we immediately obtain our thesis. $\blacksquare$

{\bf Theorem 3.} Suppose that Alice and Bob share an arbitrary maximally entangled state $|\Phi\rangle_{AB}$. Alice performs a measurement in one of two bases ${\cal A}^{(s)}=\{|a_{k}^{(s)}\rangle \}$ corresponding to one-dimensional projectors $\{P_{k}^{(s)}=|a_{k}^{(s)}\rangle \langle a_{k}^{(s)}|\}$ and Bob performs a measurement in one of two bases ${\cal B}^{(s)}=\{|b_{j}^{(s)}\rangle \}$  corresponding to one-dimensional projectors $\{Q_{j}^{(s)}=|b_{j}^{(s)}\rangle \langle b_{j}^{(s)}|\}$. The following uncertainty relation holds:
\begin{eqnarray}
I(A^{(1)}:B^{(1)})+I(A^{(2)}:B^{(2)}) \leq 2 \log d + \log c'
\label{eq:theorem-3}
\end{eqnarray}
where $d$ is dimension of each party's Hilbert space and and $c'=\max_{V} \max_{i,j} |\langle b_{i}^{(1)}|VU^TV^{\dagger}|b_{j}^{(2)} \rangle|^{2}$ with  $U$ chosen in such a way that relation
\begin{eqnarray}
U^{\dagger}|a_{k}^{(2)}\rangle=|a_{k}^{(1)}\rangle
\end{eqnarray}
is satisfied for all $k$ and maximum taken over all unitary operations $V$.

{\it Proof.} Proof immediately follows from Theorem 2, as maximization over $V$ gives the upper bound in the worst case (i.e. for a maximally entangled state for which the sum of two mutual informations is maximal). $\blacksquare$

There is still an open question if the uncertainty relation (\ref{eq:theorem-3}) holds for nonmaximally entangled states. Let us suppose that mutual information between the results of measurements performed on the nonmaximally entangled state of the form $|\Phi\rangle_{AB}=\sum_{i}\sqrt{\lambda_{i}}|\lambda_{Ai}\rangle \otimes |\lambda_{Bi}\rangle$ ($|\lambda_{Ai}\rangle \otimes |\lambda_{Bi}\rangle$ is Schmidt basis) is smaller than mutual information between results of measurements performed on the maximally entangled state $|\Phi\rangle_{AB}=\frac{1}{\sqrt{d}}\sum_{i}|\lambda_{Ai}\rangle \otimes |\lambda_{Bi}\rangle$. In such a case we can replace the former by the latter and prove analog of Theorems 2 and 3 for the results of measurements performed on nonmaximally entangled state. However, for a measurement of the observables $A=X+Z$  on Alice's side and $B=X-Z$ on Bob's side mutual information is $I(A:B)=0.049$
when the parties perform the measurement on the nonmaximally entangled state
$|\Phi\rangle_{AB}=\sqrt{0.0332}|0\rangle \otimes |0\rangle+\sqrt{0.9668}|1\rangle \otimes |1\rangle$, and it is $I(A:B)=0$ when the parties perform the measurement on the  maximally entangled state $|\Phi^+\rangle_{AB}$. Hence, the straightforward generalization of the proof (but not the uncertainty relation) fails.

\section{Conclusions}
We have proposed mutual uncertainty relations within distant labs paradigm
which bound the sum of mutual informations  between Alice's and Bob's results
of measurements for different observables. We have proved these uncertainty relations for some states and observables. We have also tested numerically the inequalities and found
no numerical violations. Remarkably, the mutual uncertainty relation
(\ref{eq:4}) for \ovt\ observables (one on Alice's side and two on Bob's side) is much stronger than Hall uncertainty relation (\ref{eq:3}) derived from the Maassen-Uffink uncertainty relation.
On the other hand the uncertainty relation for \tvt\ observables
has the coefficient $c'$ on the RHS analogous to the coefficient $a$ on the RHS of Maassen-Uffink uncertainty relation. It would be interesting to check if the following uncertainty relation holds
\begin{eqnarray}\label{eq:conc}
I(A^{(1)}:B^{(1)})+I(A^{(2)}:B^{(2)}) \leq 2 \log d + \log c''',
\end{eqnarray}
where $c'''=\max_V \sum |\langle b_{i}^{(1)}|VU^TV^{\dagger}|b_{j}^{(2)} \rangle|^{2}$. Here $U$ is chosen in such a way that relation $U^{\dagger}|a_{k}^{(2)}\rangle=|a_{k}^{(1)}\rangle$ is satisfied for all $k$ and $V$ is unitary operator. The sum is taken over $d$ largest coefficients $|\langle b_{i}^{(1)}|VU^TV^{\dagger}|b_{j}^{(2)} \rangle|^{2}$.

Let us also make a remark on generalization of our uncertainty relations to continuous variables when both parties measure operators such as position $x$ and momentum $p$. In such a case there always exists a
state for which at least one mutual information can be arbitrarily large (it is related to the fact that the Hilbert space is infinite dimensional) and hence one cannot bound the sum of both mutual informations.
In order to obtain non-trivial uncertainty relations for continuous variables one should encompass the finite resources such as bounded average energy which would make the relations quite different from the proposed ones.

{\it Note added.} After submission of this paper our conjectured uncertainty relation for one vs. two observables was proved in general case by P. J. Coles and M. Piani \cite{Coles1}.

\section{Acknowledgements}
We thank Otfried G\"{u}hne, Karol Horodecki, Adam Miranowicz and Renato Renner for valuable discussions. We also thank Michael J. W. Hall and Karol \.{Z}yczkowski for helpful comments. This work
is supported by the ERC Advanced Grant QOLAPS and National Science Centre project Maestro DEC-2011/02/A/ST2/00305.

\section{Appendix}

Here we present details of numerical calculations. The numerical evidence was obtained using Genetic Algorithm.  Genetic
organisms were represented as vectors of real numbers in the range 0 to 1
which were subjected to mutation, crossover and selection.  The
crossover was done by random selection of elements which are being
swapped (i.e., individual elements were selected for swapping).  We
used a population of 25 organisms and the elite of three best organisms
were always taken to the next generation unchanged.

We used a technique which we call \emph{mutation scaling}, which is
supposed to allow to approach the (maybe local)
maximum.  The mutation was done by adding $s (1 - 2r)$ to the elements
selected for mutation, where $r$ is a random number in the range 0 to 1
while $s$ is a scaling factor. The scaling factor starts with 1 and is
increased to $s' = \min(1, 1.1s)$ if the current generation brings
improvement (i.e., in the current generation there is an organism which
is better than the best organism in the previous generation). The
scaling factor is decreased to $s'= s / 1.05$ if the current generation
does not bring improvement (i.e., the best organism in the current
generation is the same as the best organism in the previous generation).
If $s$ decreases to $10^{-9}$ the scaling factor is set back to 1.

In order to optimize over the unitary operators of a given dimension we used a
function which returns a unitary matrix given a vector of random numbers
(implementation of an algorithm proposed in \cite{PZK98bb-unitary})
suitable in optimizations with Genetic Algorithm.

\end{document}